\documentclass[aip,amsmath,amssymb,reprint,longbibliography]{revtex4-2}

\usepackage[colorlinks,citecolor=blue,linkcolor=red,urlcolor=blue]{hyperref}
\usepackage{color}
\usepackage{graphicx}
\usepackage{subfigure}
\usepackage{verbatim}
\usepackage{fourier}

\usepackage{hyperref}
\usepackage{graphicx}
\usepackage{dcolumn}
\usepackage{bm}
\usepackage{xcolor} 
\usepackage{orcidlink}
\definecolor{color1}{rgb}{0,0.25,0.70}

\begin{document}

\preprint{APS/123-QED}

\title{Dynamic charges effect on infrared
	dielectric response of polar materials}

\author{Wei-Zhe Yuan~\orcidlink{0009-0001-4678-4344}}
\affiliation{School of Energy Science and Engineering, Harbin Institute of Technology, Harbin 150001, China}
\affiliation{Key Laboratory of Aerospace Thermophysics, Ministry of Industry and Information Technology, Harbin 150001, China}
\author{Yangyu Guo~\orcidlink{0000-0003-2862-896X}}
\email[]{yyguo@hit.edu.cn}
\affiliation{School of Energy Science and Engineering, Harbin Institute of Technology, Harbin 150001, China}
\affiliation{Key Laboratory of Aerospace Thermophysics, Ministry of Industry and Information Technology, Harbin 150001, China}
\author{Hong-Liang Yi~\orcidlink{0000-0002-5244-7117}}
\email[]{yihongliang@hit.edu.cn}
\affiliation{School of Energy Science and Engineering, Harbin Institute of Technology, Harbin 150001, China}
\affiliation{Key Laboratory of Aerospace Thermophysics, Ministry of Industry and Information Technology, Harbin 150001, China}
\date{\today}

\begin{abstract}
Predictive modeling of the infrared dielectric function in polar materials is crucial for thermal management and infrared devices design. While the Green-Kubo molecular dynamics (MD) framework provides a nonperturbative route to compute dielectric responses from dipole fluctuations, yet it commonly relies on the fixed-charge approximation that neglects dynamic charge redistribution during atomic motion. Here, we employ a machine-learning neuroevolution potential with dynamic charges (qNEP) combined with Green-Kubo MD to investigate the dynamic charge effect on the infrared dielectric response of rutile TiO$_2$, a material with large Born effective charges (BEC). Our results show that dynamic charge effects become increasingly important at elevated temperatures and are essential for accurately predicting longitudinal optical phonon features and infrared reflectance. This work establishes that accurate prediction of infrared optical properties in polar materials under thermal excitation requires explicit treatment of dynamic charge evolution.
\end{abstract}

\maketitle

\section{Introduction}
The infrared dielectric function is a key physical quantity governing light-matter interactions. Accurately understanding how it depends on frequency and temperature is essential for designing and optimizing infrared optical devices, thermal management materials, and energy conversion systems~\cite{2022JAP,howell2020thermal,modest2021radiative,howell2023thermal,Zhao_2024,CHENG2025253,2019-nanophoto,2015nanoph}. Under extreme thermal conditions, in particular, the complex interplay between ionic motion and electron‑cloud response introduces pronounced anharmonic and dynamic features into the dielectric response~\cite{Tong2020PRB,Zhou2024PRL}, posing significant challenges for theoretical methods that aim to predict the dielectric function reliably.

Although the Lorentz model with first-principles perturbative theory relate the infrared dielectric response to lattice dynamics~\cite{BAO20121683,Tong2020PRB}, their underlying quasiparticle approximation breaks down at elevated temperatures~\cite{PhysRevX.12.041011,PhysRevB.109.014310}, representing a fundamental limitation. In contrast, Green-Kubo MD simulation provides a non-perturbative framework that directly extracts the infrared dielectric function from equilibrium dipole moment fluctuations~\cite{PhysRev.123.777,Gangemi_2015,DOMINGUES2018220}. This approach naturally incorporates anharmonic interactions to all orders and is applicable to large systems with complex structures. Combined with machine-learning potentials, it has successfully reproduced the infrared dielectric properties of bulk polar materials~\cite{Chen2021JAP,Yuanwz2025PRB}.

However, most current Green-Kubo MD implementations rely on a key approximation: the dipole moment is computed using fixed atomic partial charges~\cite{PhysRev.123.777,Gangemi_2015,DOMINGUES2018220,Chen2021JAP,Yuanwz2025PRB}. This fixed-charges approximation presumes that the charge distribution remains unchanged under instantaneous atomic configurations~\cite{chemrev,PhysRevB.106.L180303,Zhong2025,2025JAP_Luo}. For simple systems with relatively uniform chemical environments and narrow charge distributions~\cite{deng2019electrostatic}, this approximation shall be acceptable~\cite{Yuanwz2025PRB}. Yet for materials that exhibit sizable partial charges~\cite{PhysRevB.109.205204}, significant charge transfer~\cite{PhysRevB.58.6224,PhysRevB.110.L041101}, or structural disorder~\cite{PhysRevLett.79.1766,PhysRevMaterials.7.045604}, a static and fixed charge assignment cannot capture the real-time contribution of electron cloud redistribution during atomic motion. Consequently, the total polarization fluctuations are inaccurately estimated, which finally affects the predicted infrared dielectric function. Moving beyond the static fixed-charges approximation and self-consistently accounting for the dynamic evolution of charge distributions in atomistic simulations is therefore an essential step toward improving the predictive reliability of infrared dielectric response.

\begin{figure}[htbp]
	\centering
	\includegraphics[width=1\linewidth]{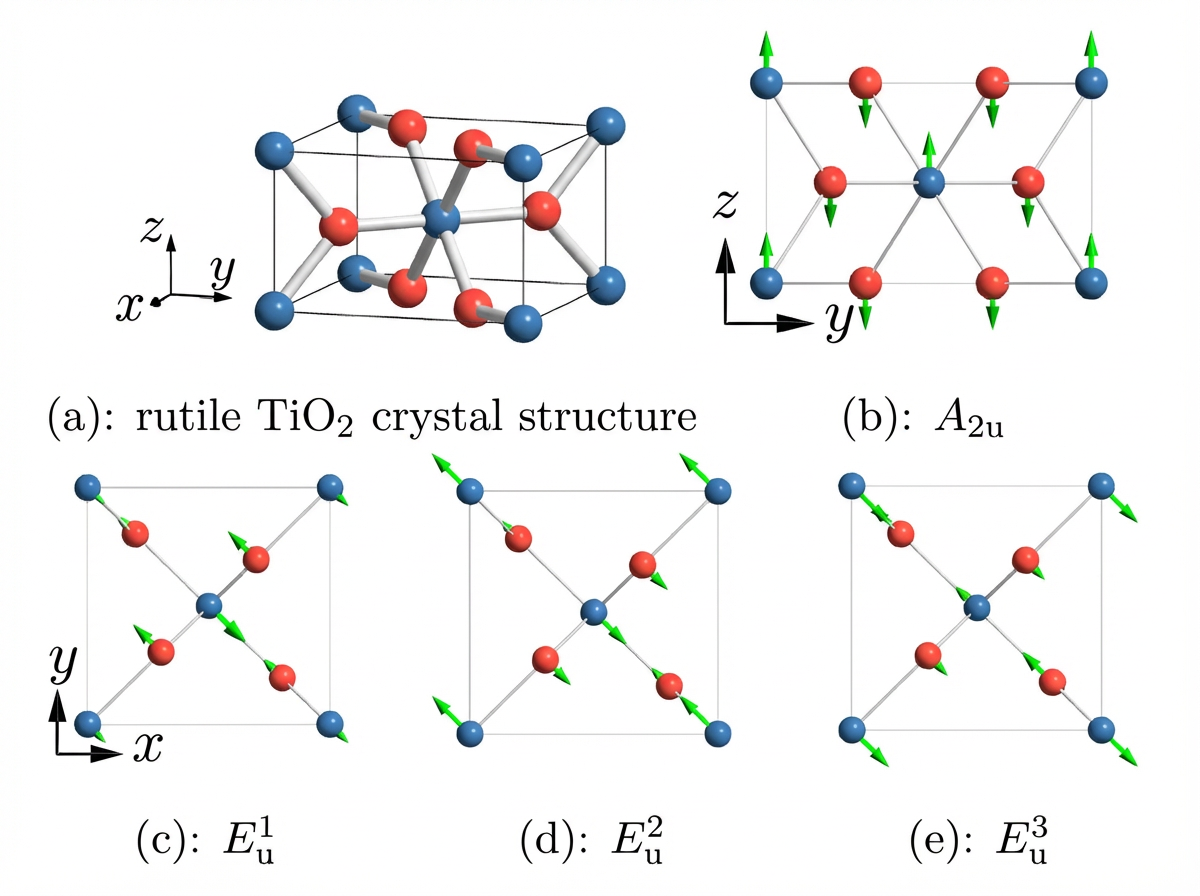}
	\caption{(a) Crystal structure of rutile TiO$_2$ (Ti: red atoms, O: blue atoms). (b)-(e) Schematic views
		of atomic eigenvector for the $A_{2u}$ mode and the three $E_u$ modes.}
	\label{fig:tio2}
\end{figure}

In this work, we employ atomistic simulations based on the neuroevolution potential with dynamic charges (qNEP)~\cite{qnep} to investigate the role of charge fluctuations on the infrared dielectric response of rutile TiO$_2$ (Fig.~\ref{fig:tio2}), a prototypical polar material  with large Born effective charges (BEC)~\cite{PhysRevB.49.14730}. We demonstrate that the fixed-charge approximation significantly overestimates both the longitudinal optical (LO) phonon frequencies and the infrared reflectance, whereas the qNEP model with dynamic charges achieves much better agreement with experimental data across a wide temperature range. Furthermore, we reveal the temperature dependence of the BEC distribution, showing that with increasing temperature the average BECs decrease while their fluctuations broaden, a behavior that cannot be captured by fixed-charge models. Our study thus demonstrates the critical importance of dynamic charge effects for the accurate prediction of infrared properties in polar materials at finite temperatures.

The remaining of this work is organized as follows: The
methodology and simulation detals will be introduced in Sec.~\ref{sec2}, followed by a discussion of the results in Sec.~\ref{sec3}, and the concluding remarks will be made in Sec.~\ref{sec4}.
\section{METHODOLOGY}\label{sec2}
In this section, we systematically outline the methodology used to predict the infrared optical properties of polar materials. We begin by introducing  Green-Kubo formalism in Sec.~\ref{sec2A}. Subsequently, we provide the training procedure for MLP in Sec.~\ref{sec2B}, as well as the details of MD simulation in Sec.~\ref{sec2C}. 
\subsection{Green-Kubo formalism}\label{sec2A}
The local dielectric function is related to the dielectric
susceptibility $\chi_{\alpha \beta}(\omega)$ as ~\cite{PhysRev.123.777,Gangemi_2015},
\begin{equation}
	\varepsilon _{\alpha \beta}(\omega )=\delta _{\alpha \beta}+\chi _{\alpha \beta}\left( \omega \right) +\varepsilon _{\infty ,\alpha \beta}-1,
\end{equation}
where $\omega$ is the photon frequency, $\delta_{\alpha\beta}$ is the Kronecker delta, and the imaginary part of susceptibility $\chi''_{\alpha \beta}$ is calculated via the fluctuation-dissipation theorem (i.e. Green-Kubo formalism)~\cite{PhysRev.123.777,Gangemi_2015,DOMINGUES2018220,Chen2021JAP,Yuanwz2025PRB}:
	\begin{equation}\label{eq:chi}
	\begin{aligned}		
		\chi''_{\alpha \beta}\left( \omega \right)=&\mathrm{Im}\frac{1}{\varepsilon _0Vk_{\mathrm{B}}T\omega ^2}\bigg[ \left< J_{\alpha}\left( 0 \right) \cdot J_{\beta}\left( 0 \right) \right> 
		\\
		&+i\omega \int_0^{\infty}{\left< J_{\alpha}\left( 0 \right) \cdot J_{\beta}\left( t \right) \right> e^{i\omega t}dt} \bigg],
	\end{aligned}	
\end{equation}
where $V$ is the system volume, with $\varepsilon_0$ the dielectric permittivity of the vacuum and $k_{\mathrm{B}}$ represents the Boltzmann constant. In Eq.(\ref{eq:chi}), the total ionic polarization of the system is calculated as the current density $\mathbf{J}\left( t \right)$:
\begin{equation}
	\mathbf{J}\left( t \right) =\sum_{lb}{\mathbf{Z}_{lb}\left( t \right) \otimes \mathbf{v}_{lb}\left( t \right) ,}
\end{equation}
where $\mathbf{v}_{lb}$ the atomic velocity and $\mathbf{Z}_{lb}$ is the dynamic atomic BEC tensor. For centrosymmetric crystals such as rutile TiO$_2$, intrinsic atomic dipole moments vanish by symmetry, and these intrinsic contributions primarily affect the dielectric response at low frequencies~\cite{10.1063/1.3432620}. The real part of susceptibility $\chi'_{\alpha \beta}$ is then calculated using Kramer-Kronig relation~\cite{PhysRev.104.1760}
\begin{equation}
	\chi '_{\alpha \beta}\left( \omega \right) =\frac{2}{\pi}\int_0^{\infty}{\frac{\omega '\chi ''_{\alpha \beta}\left( \omega ' \right)}{\omega '^2-\omega ^2}d\omega '}.
\end{equation}
As high-frequency dielectric constant $\varepsilon _{\infty ,\alpha \beta}$ is well known to be overestimated in density functional theory (DFT), we use the experimental valus of $\varepsilon _{\infty ,xx}=\varepsilon _{\infty ,yy}=6$ and  $\varepsilon _{\infty ,zz}=7.8$~\cite{1974PRB_TiO2} in the following calculations. The influence of $\varepsilon_\infty$ on reflectance is discussed in Fig.~\ref{fig:rt4} of Appendix.~\ref{sec:a4}. With the predicted dielectric function of rutile TiO$_2$, the reflectance normal to the surface is given by
\begin{equation}
	R_{\alpha \alpha}=\left| \frac{\sqrt{\varepsilon _{\alpha \alpha}\left( \omega \right)}-1}{\sqrt{\varepsilon _{\alpha \alpha}\left( \omega \right)}+1} \right|^2.
\end{equation}
\subsection{Machine learning potentials}\label{sec2B}
We use the qNEP~\cite{qnep} to construct accurate machine learning potential for rutile TiO$_2$. The original NEP method does not include electrostatic interactions,  which is instead incorporated in qNEP. More importantly, in qNEP, partial charges are treated as latent features of the model and are determined implicitly by fitting the sum of the electrostatic and short-range NEP contributions to the total target energies and forces~\cite{qnep}. Subsequently, the partial charges are scaled to BEC according to $\varepsilon_\infty$ trained from reference charge data~\cite{qnep} according to Eq.~\ref{eq:scale1} and \ref{eq:scale2}, 
\begin{equation}\label{eq:scale1}
\tilde{q}_i=\sqrt{\varepsilon _{\infty}}q_i,
\end{equation}
\begin{equation}\label{eq:scale2}
Z_{i\alpha \beta}=\frac{\partial P_{\alpha}}{\partial r_{i\beta}}=\tilde{q}_i\delta _{\alpha \beta}+\sum_j{r_{j\alpha}\frac{\partial \tilde{q}_j}{\partial r_{i\beta}}},
\end{equation}
$q_i$ and $\tilde{q}_i$ are the partial and scaled charges, respectively. This allows the model to capture electrostatic effects without requiring predefined charge assignments, which is the key difference from the standard NEP framework.
We consider mode 1~\cite{qnep} for evaluating the electrostatic energy where both the real-space and reciprocal-space contributions
are included according to Eq.~\ref{eq:U_ES}~\cite{qnep},
\begin{equation}\label{eq:U_ES}
	U^{\mathrm{ES}}=U^{\mathrm{r}}+U^k+U^{\mathrm{s}},
\end{equation}
where $U^{\mathrm{r}}$ is the real-space component, $U^k$ is the
reciprocal-space component, and $U^{\mathrm{s}}$ is the self-energy. The total energy in qNEP is given by the sum of the NEP energy and the electrostatic energy,
\begin{equation}
	U^{\mathrm{tot}}=U^{\mathrm{NEP}}+U^{\mathrm{ES}}.
\end{equation}
A total of 797 structures were sampled from two types of configurations: those with random atomic displacements of $0.1\ \text{\AA}$ and those generated via MD simulations.  Among the full set of training structures, only a subset (specifically 150 structures with 64 atoms) includes BEC reference data. The remaining structures do not have such data, as this number is sufficient to obtain converged high-frequency dielectric constants~\cite{kim2025longrangeelectrostaticsmachinelearning}. This treatment will also reduce the computational cost since BEC calculations are computationally much more demanding than calculations of energies, forces, or virials. The \textsc{Quantum Espresso} package~\cite{2020QE} with the projector-augmented wave method is used to obtain the energy, forces, and virial of each structure. In the DFT calculations, the local density approximation (LDA) functional is used to describe the exchange-correlation of electrons~\cite{PhysRevB.23.5048}. The kinetic energy cutoff for wave function and charge density have been set to 52 Ry and 576 Ry, respectively. Convergence threshold for self consistency was $10^{-6}$ eV. The $\mathbf{k}$-point mesh was set to Gamma only and $2\times2\times2$ for structures with 216 atoms and 64 atoms respecttively. Then, 80\% and 20\% structures were randomly selected to form the training and testing datasets, respectively. 

\begin{figure}[htbp]
	\centering
	\includegraphics[width=1\linewidth]{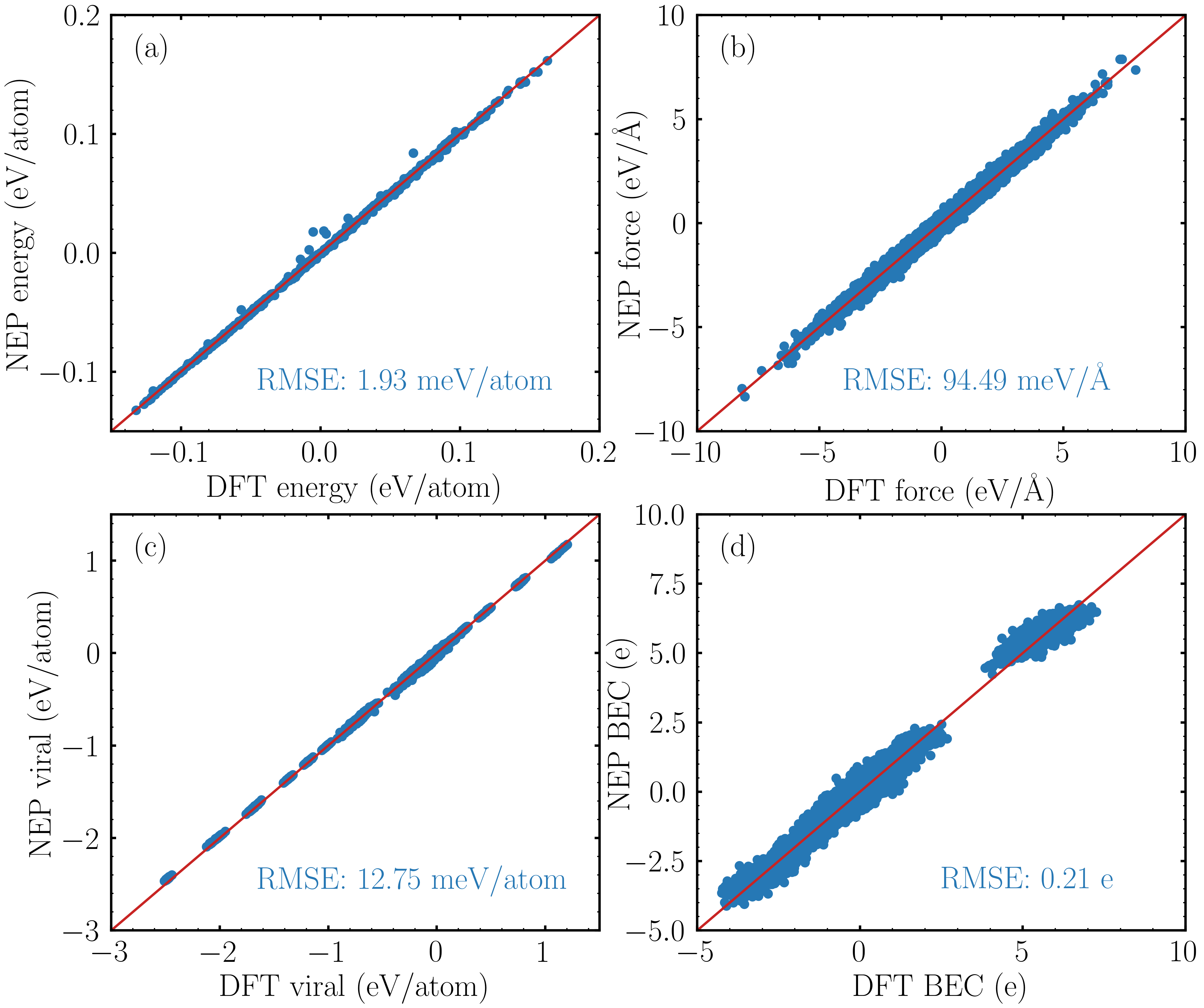}
	\caption{Accuracy evaluation of the qNEP of rutile TiO$_2$: The partial plots of (a) total energy, (b) atomic forces, (c) virial for NEP training and (d) Born effective charges (BEC). The insets show the root mean square error (RMSE) of the training datasets.}
	\label{fig:train}
\end{figure}	
During the training processes, the radial and angular descriptor components were set to a cutoff radius of 5~\AA.  The parity plots and accuracy metrics are shown in Fig.~\ref{fig:train}. The root mean square error (RMSE) values of the total
energy, atomic forces, and virial for the testing dataset are: 1.93 meV/atom, 94.49 meV/\AA~and 12.75 meV/atom respectively. The RMSE demonstrate some improvement in the accuracy of forces, and stresses for the qNEP models compared to the NEP model (forces: 124.30 meV/\AA~and virial: 19.69 meV/atom), highlighting the importance of long-range electrostatic interactions in polar materials. 
To further validate the qNEP accuracy, we computed the phonon dispersion using
spectral energy density analysis as shown in Fig.~\ref{fig:2ph} of Appendix~\ref{sec:a1}.
\subsection{MD simulation}\label{sec2C}
The Green-Kubo calculation is done based on equilibrium molecular dynamics~(EMD) as implemented in Graphics Processing Units Molecular Dynamics~(\textsc{GPUMD}) package~\cite{GPUMD4} with qNEP.  The particle-particle
particle-mesh method~\cite{EASTWOOD1980215} is implemented for the treatment of long-range Coulomb interaction. During the EMD simulation, a supercell  with $10\times10\times10$ primitive cells and a time step of 0.5 fs are adopted. The size of supercell has been tested to be sufficiently capture the long-range interaction well. In addition, periodic boundary conditions are imposed along the three directions of the system. First, $1\times10^{6}$ time steps are run under the \textit{NPT}~(isothermal-isobaric) ensemble for structure relaxation. Then, another $1\times10^{6}$ time steps are run under the \textit{NVE}~(microcanonical) ensemble. The trajectory and dynamic BEC tensor of the system are output once per 10 time steps during the \textit{NVE} run. Five independent
EMD simulations are conducted to reduce the statistical fluctuations.
\begin{figure*}[ht]
	\centering
	\includegraphics[width=0.85\linewidth]{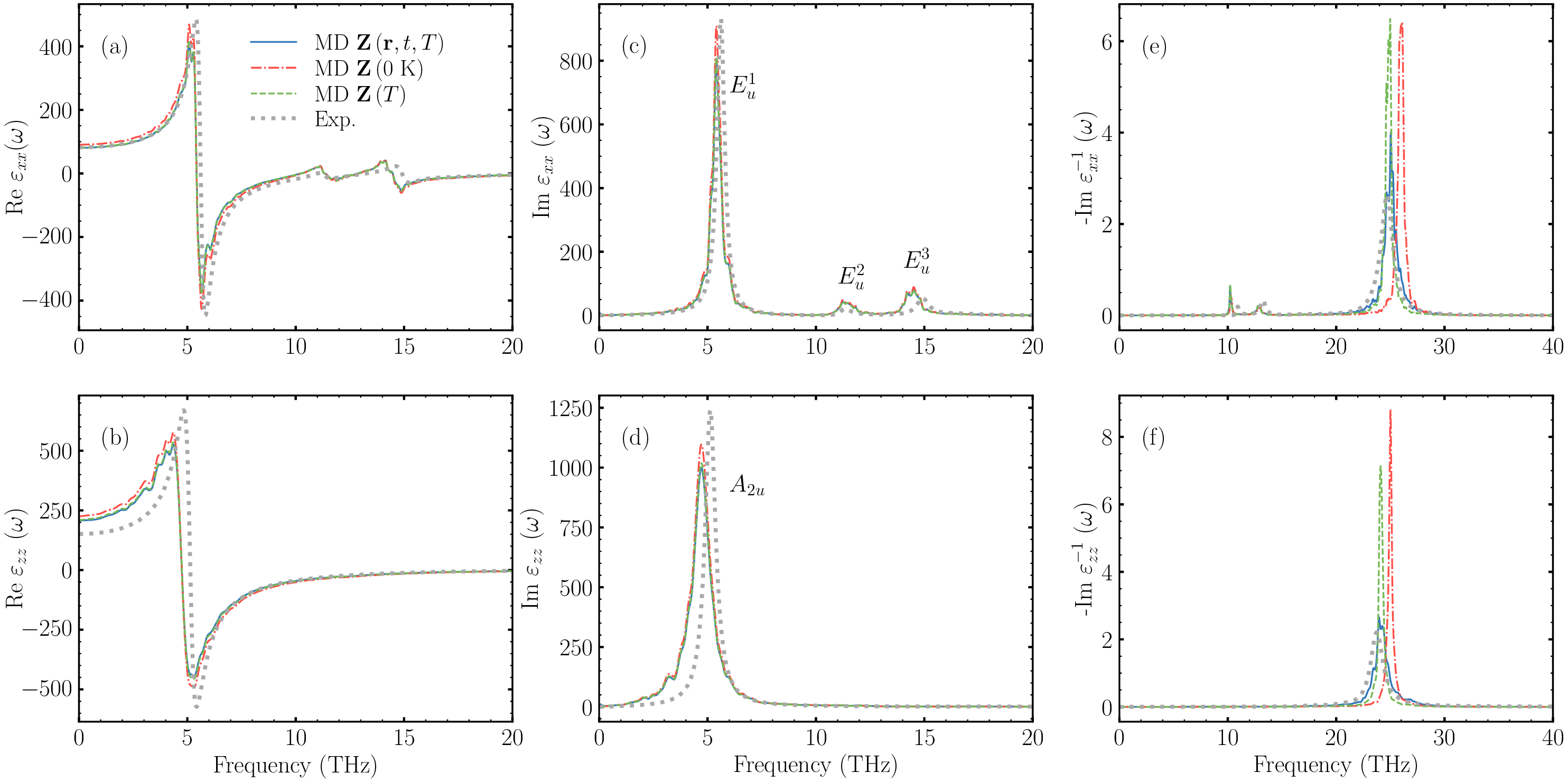}
	\caption{Infrared dielectric response of rutile TiO$_2$ at 295 K along $x$ and $z$ directions, obtained via MD with dynamic charge [$\mathbf{Z}\left( \mathbf{r},t,T \right)$, blue solid lines], fixed charge at ground state [$\mathbf{Z}\left(0~\mathrm{K} \right)$, red dashdot lines], fixed temperature dependent charge [$\mathbf{Z}\left(T \right)$, green dashed lines], and measurement~\cite{1974PRB_TiO2} (grey dotted lines). (a),(b) Real part of dielectric function;
		(c),(d) imaginary part;
		(e),(f) inverse dielectric loss function.}
	\label{fig:eps}
\end{figure*}

\section{RESULTS AND DISCUSSIONS}\label{sec3}
First, we computed the infrared dielectric function at room temperature (295 K) and compared it with experimental data~\cite{1974PRB_TiO2} to assess the accuracy of the qNEP approach, as shown in Fig. \ref{fig:eps}. MD simulations with dynamic charges [$\mathbf{Z}\left( \mathbf{r},t,T \right)$] capture the main spectral features of the measured dielectric function as shown in Figs.~\ref{fig:eps}(a)-(d). Notably, qNEP accurately predicts the imaginary part of inverse dielectric function compared to the experiment as seen in Figs.~\ref{fig:eps}(e) and (f). Since the poles of the dielectric loss function $\varepsilon^{-1}$ correspond to $\omega_{\mathrm{LO}}$, which are governed by long-range electrostatic interactions~\cite{Schubert2004}. Our results demonstrate qNEP's ability to capture these long-range electrostatic interactions. 
Some discrepancies between the theoretical and experimental results in Figs.~\ref{fig:eps}(b) and (d) mainly originate from a slight underestimation of the TO phonon frequency along the $z$-direction. This deviation is associated with the choice of pseudopotential used in the simulations, as also discussed in Ref.~\cite{TiO2_2023PRB}.

\begin{figure}[htbp]
	\centering
	\includegraphics[width=1\linewidth]{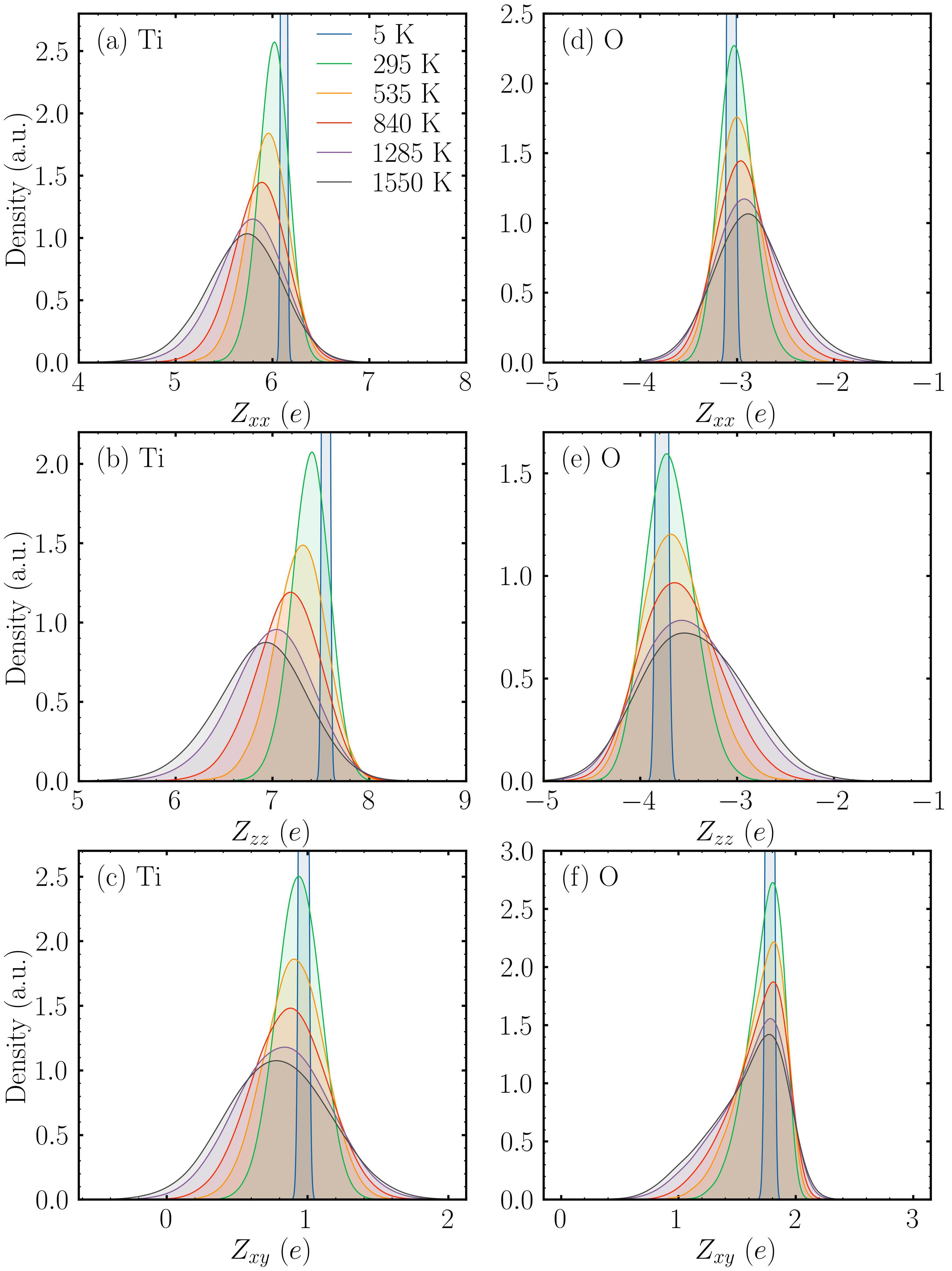}
	\caption{Temperature-dependent Born effective charge distribution of rutile TiO$_2$ predicted from qNEP. Panels (a)-(c) show the charge  distributions of Ti atoms for
		$Z_{xx}$, $Z_{zz}$, and $Z_{xy}$, respectively,
		while panels (d)-(f) show the corresponding charge
		distributions of O atoms.}
	\label{fig:charget}
\end{figure}

For MD simulations using NEP with fixed charges [$\mathbf{Z}\left(0~\mathrm{K} \right)$] derived from DFPT (Ti: $Z_{xx}=6.36,Z_{xy}=\pm1.00,Z_{zz}=7.66$; O: $Z_{xx}=-3.18,Z_{xy}=\pm1.81,Z_{zz}=-3.83$), the predicted real and imaginary parts of the dielectric function are nearly identical to those obtained with qNEP. However, the imaginary part of $\varepsilon^{-1}$ deviates significantly from both qNEP and the experiment, especially overestimating the $\omega_{\mathrm{LO}}$. 
We attribute this discrepancy partly to the slight reduction of the average BEC from the ground state to 295 K (explicitly shown in Fig. \ref{fig:charget} later),
Nevertheless, even accounting for this effect [$\mathbf{Z}\left(T \right)$], the fixed charges model only accurately predicted the $\omega_{\mathrm{LO}}$, while yields an unrealistically narrow linewidth for the LO phonon that might lead to inaccurate predictions of the reflectance as shown below.
This indicates that the linewidth of $\varepsilon^{-1}$ for these materials with large BEC is not only determined by phonon scattering, but also affected by the dynamic charges effect.

\begin{figure}[htbp]
	\centering
	\includegraphics[width=1\linewidth]{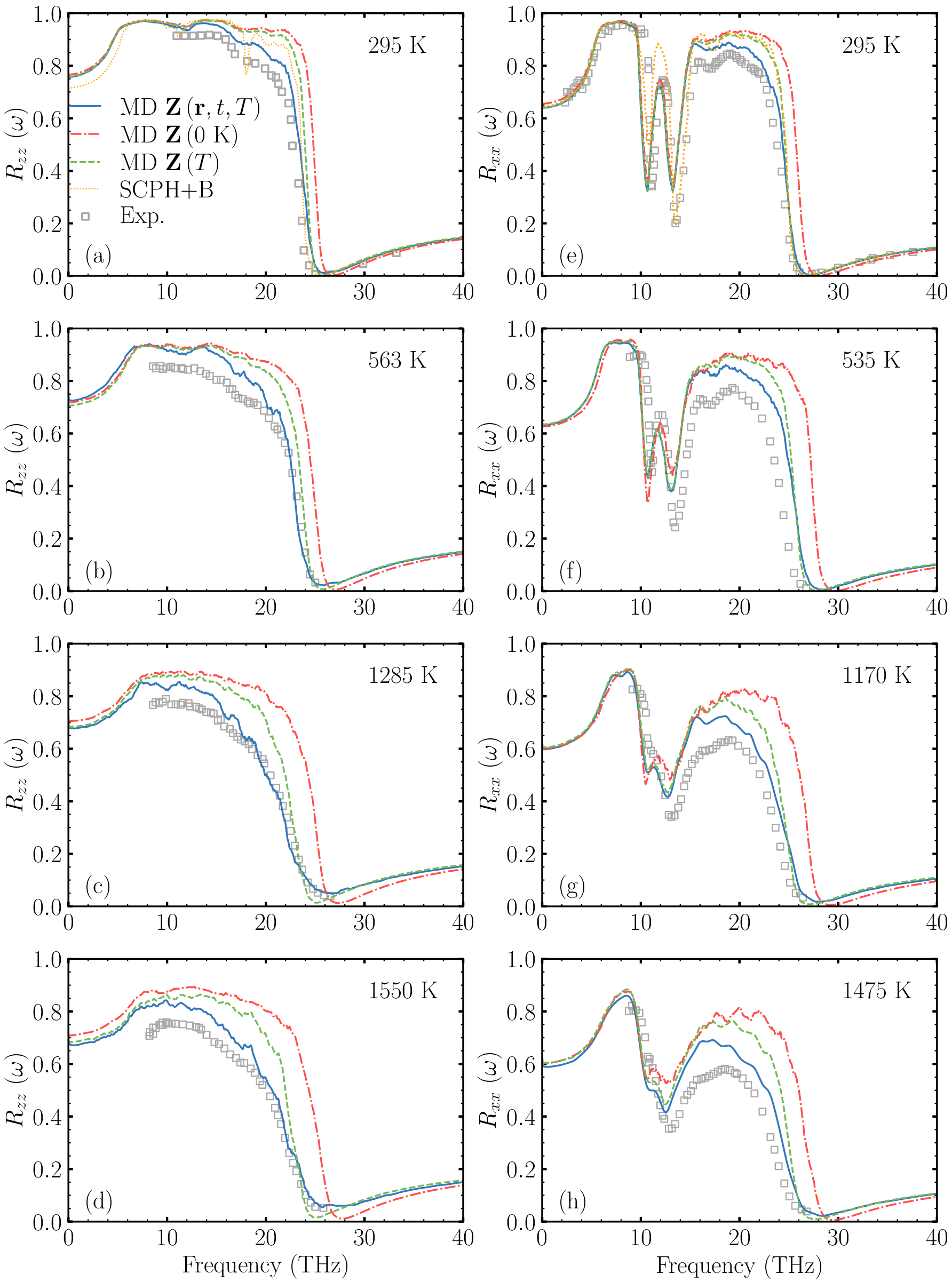}
	\caption{The temperature-dependent infrared reflectance of rutile TiO$_2$ in $z$ [(a)-(d)] and $x$ directions [(e)-(h)], obtained via MD with dynamic charge [$\mathbf{Z}\left( \mathbf{r},t,T \right)$, blue solid lines], fixed charge at ground state [$\mathbf{Z}\left(0~\mathrm{K} \right)$, red dashdot lines], fixed temperature dependent charge [$\mathbf{Z}\left(T \right)$, green dashed lines], SCPH+B~\cite{TiO2_2023PRB} (orange dotted lines) and measurement~\cite{1974PRB_TiO2} (grey squares).}
	\label{fig:rt}
\end{figure}

For the reflectance of rutile TiO$_2$ at 295 K, qNEP  with dynamic charges provides superior agreement with experimental measurements compared to fixed-charge approaches, as illustrated in Figs.~\ref{fig:rt}(a) and (e). In particular, simulations using the NEP potential with DFPT-derived static charges overestimate both the reflectance and the position of sharp decrease (corresponding to $\omega_{\mathrm{LO}}$). Prediction employing the fixed mean BEC at 295 K, while correctly capturing the position of sharp decrease, still overestimates the reflectance due to the underestimation of linewidth of LO phonon. This overestimation aligns with previous perturbative calculations based on self-consistent phonon theory with bubble self-energy (SCPH+B) that also employed fixed charges~\cite{TiO2_2023PRB}. Hence, the discrepancy might originate from charge fluctuations in the instantaneous atomic environment.

With rising temperature, the average BEC decreases and its distribution broadens, as demonstrated in Fig.~\ref{fig:charget}. A previous study~\cite{GERVAIS1976191} also pointed out that the average value of the effective charge of rutile-type TiO$_2$ would decrease with the rise in temperature based on the degree of LO-TO spiltting, but it was unable to provide the distribution of the charge. At 1550 K, BEC fluctuations reach approximately 3 $e$, rendering them poorly described by a simple Dirac-delta distribution. Consequently, dynamic charges effects are expected to become more pronounced at elevated temperatures. In Fig.~\ref{fig:rt}, we further compare the temperature-dependent reflectance computed with dynamic and fixed charges against experimental measurements~\cite{1974PRB_TiO2}. The qNEP simulations with dynamic charges show excellent agreement with experiment at all temperatures, outperforming methods that employ either temperature-dependent fixed charges $\mathbf{Z}(T)$ or ground‑state charges $\mathbf{Z}(0~\mathrm{K})$. A slight overestimation of the low-frequency reflectance along the $z$ direction is likely caused by the pseudopotential, which induces an underestimation of $\omega_{\mathrm{TO}}$ and an overestimation of the static dielectric constant (see Fig.~\ref{fig:e0} of Appendix~\ref{sec:a2}). As temperature increases, the difference between reflectance predicted with dynamic and fixed charges becomes more pronounced, reflecting the growing importance of the thermal broadening in the charges distribution. Even when the temperature dependence of the fixed charges is accounted for, the reflectance remains overestimated due to the underestimation of linewidth of $\varepsilon^{-1}$ as discussed earlier. The temperature-dependence of charges is potentially came from thermal expansion, anharmonic phonon-phonon interaction and electron-phonon coupling effects~\cite{WAKAMURA1993387,PhysRevB.25.3889}. To isolate the effect of thermal expansion, we performed DFPT calculations with the quasi-harmonic approximation (QHA) as shown in Fig.~\ref{fig:bect} of Appendix~\ref{sec:a3}. The resulting temperature trend of the charges is opposite to that observed in MD simulations, and the magnitude of change is minimal. Therefore, accurately predicting infrared optical properties at finite temperatures requires explicitly capturing dynamic charges distribution.
\section{CONCLUSIONS}\label{sec4}
In summary, we have systematically investigated the influence of dynamic charge effects on the infrared dielectric response of rutile TiO$_2$ using qNEP-based Green-Kubo molecular dynamics simulations. Our results demonstrate that the conventional fixed-charge approximation, while acceptable for TO phonon predictions, fails to capture the LO phonon properties and temperature-dependent reflectance due to its neglect of charge redistribution during atomic motion. Incorporating dynamic charges significantly improves agreement with experimental infrared reflectance across a wide temperature range. The dynamic charges effect  becomes more pronounced at high temperatures. Crucially, even models employing temperature-dependent average charges still overestimate reflectance due to the underestimation of linewidth of LO phonon, highlighting the necessity of accounting for instantaneous charge fluctuations. This work establishes that accurate prediction of infrared optical properties in polar materials with large Born effective charges requires an explicit treatment of dynamic charge evolution, thereby providing a more reliable framework for the computational design of materials for thermal and infrared applications.
\begin{acknowledgments}
This paper was supported by the National Natural Science Foundation of China (Grants No. U22A20210 and 52576070), the National Natural Science Foundation for Excellent Young Scientists Fund Program (Overseas) and starting-up funding from Harbin Institute of Technology~(Grant No. AUGA2160500923).
\end{acknowledgments}
\section*{Author Declarations}
\section*{Conflict of interest}
The authors have no conflicts to disclose.
\section*{Data Availability Statement}
The data that support the findings of this study are available from
the corresponding authors upon reasonable request.
\appendix
\section{The influence of $\varepsilon_\infty$ on reflectance}\label{sec:a4}
\begin{figure}[htbp]
	\centering
	\includegraphics[width=1\linewidth]{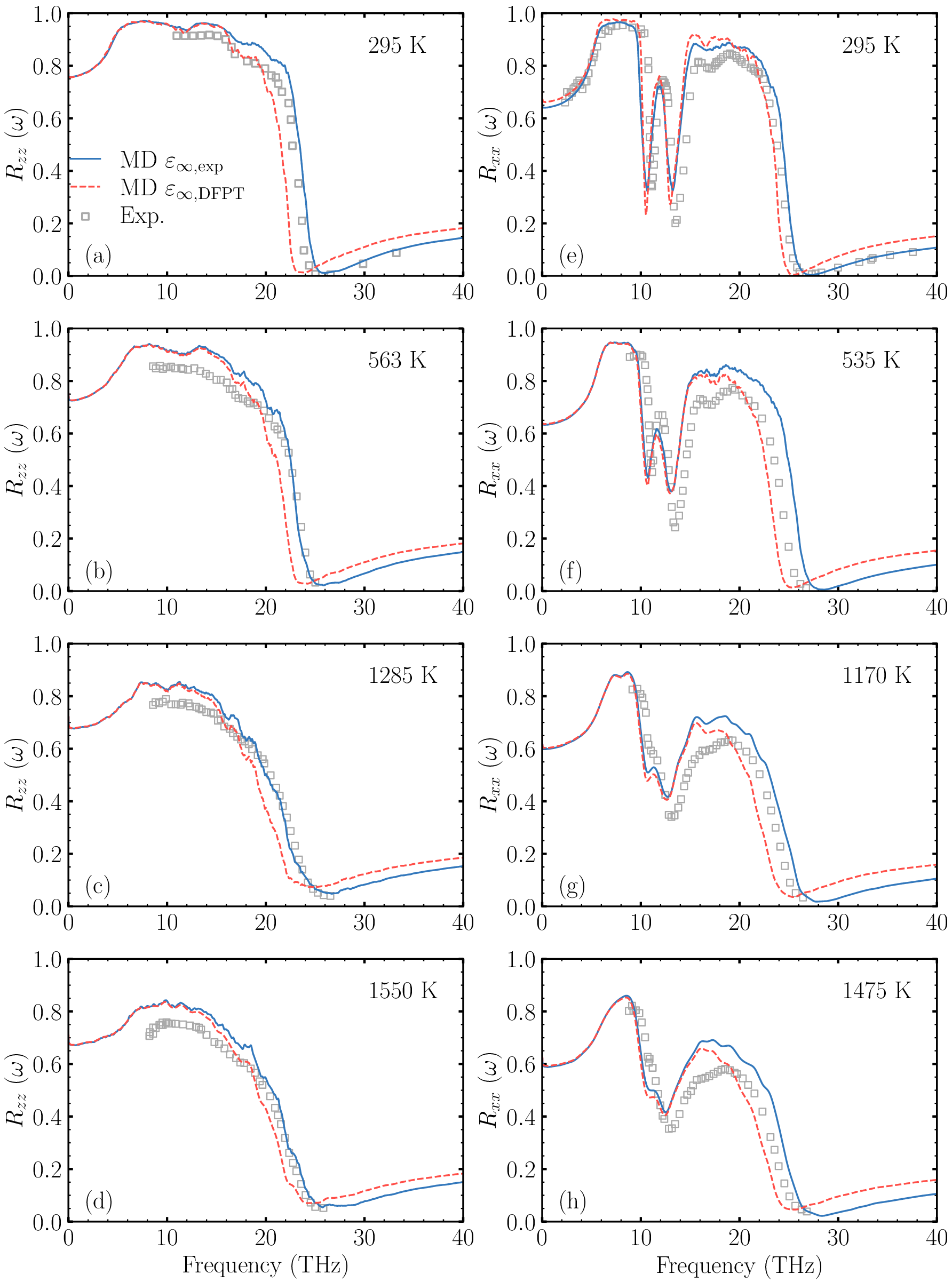}
	\caption{The temperature-dependent infrared reflectance of rutile TiO$_2$ in $z$ [(a)-(d)] and $x$ directions [(e)-(h)], obtained via qNEP with experimental (blue solid lines) and DFPT-calculated $\varepsilon_\infty$ (red dashed lines).}
	\label{fig:rt4}
\end{figure}

$\varepsilon_\infty$ calculated via DFPT
yield values of 7.55 and 8.9 for $xx$ and $zz$ directions. As shown in Fig.~\ref{fig:rt4}, the overestimation of the high-frequency dielectric constant, $\epsilon_\infty$, in DFPT calculations can cause a systematic shift in the reflectance , leading to discrepancies when comparing with measurements. This inaccuracy in DFPT typically arises from the limitations of the exchange–correlation functionals, and future improvements may achieve more reliable dielectric properties.

\section{PHONON DISPERSION RELATIONS}\label{sec:a1}
\begin{figure}[htbp]
	\centering
	\includegraphics[width=0.9\linewidth]{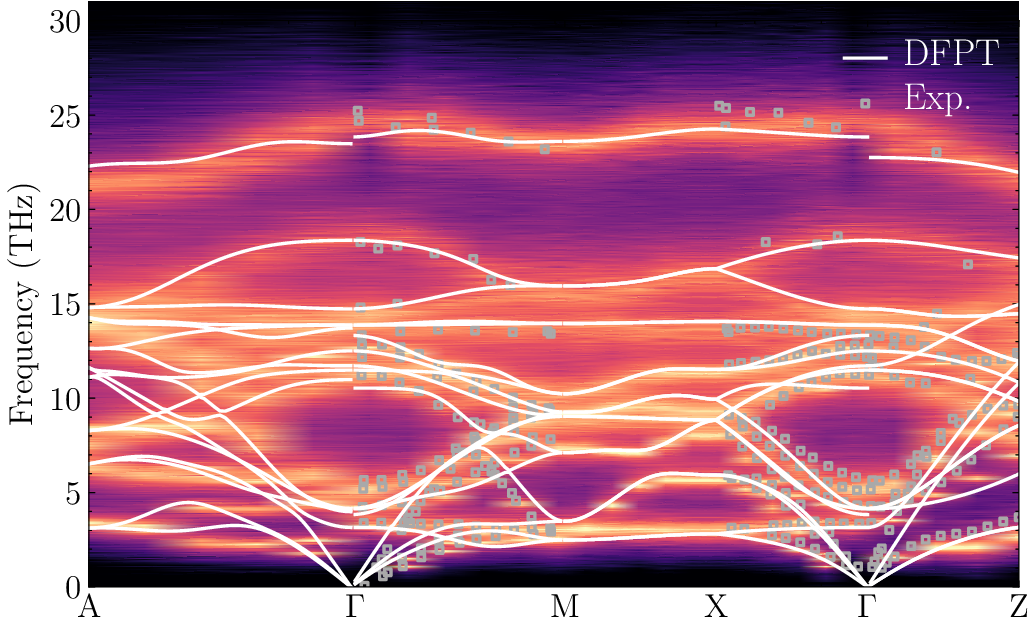}
	\caption{Phonon dispersion relations of rutile TiO$_2$ from SED (colormap) at 300 K, DFPT (white solid lines) and experiment~\cite{PhysRevB.3.3457} (grey squares).}
	\label{fig:2ph}
\end{figure}

To further validate the qNEP accuracy, we compute the phonon dispersion at 300~K through spectral
energy density analysis~\cite{PhysRevB.81.081411}
and compare it with DFPT and experiment~\cite{PhysRevB.3.3457} in Fig.~\ref{fig:2ph}. Overall, the phonon
spectra derived from qNEP show good agreement with density-functional perturbation theory (DFPT) and measurement~\cite{PhysRevB.3.3457}. In particular, it can accurately predict the longitudinal-transverse optical (LO-TO) phonon splitting near the $\Gamma$ point compared to the NEP with only short range interaction~\cite{qnep}.
\section{static dielectric constant}\label{sec:a2}
\begin{figure}[htbp]
	\centering
	\includegraphics[width=0.7\linewidth]{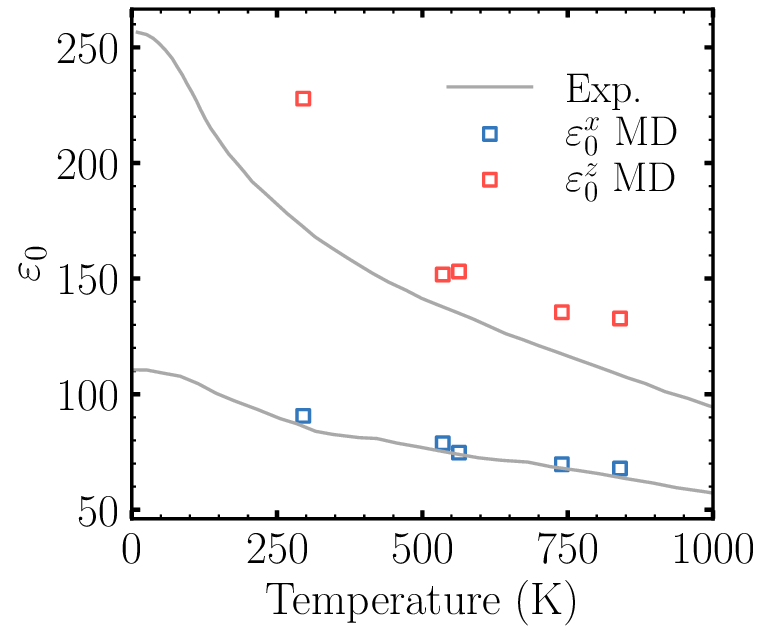}
	\caption{Temperature dependence of the static dielectric constant.
		The grey solid lines show the experimental values~\cite{PhysRev.124.1719} and the
		squares show the results from MD.}
	\label{fig:e0}
\end{figure}

Figure~\ref{fig:e0} compares the calculated temperature
dependence of $\varepsilon_{0}^{x}$ and $\varepsilon_{0}^{z}$ with the experimental data~\cite{PhysRev.124.1719}. Both $\varepsilon_{0}^{x}$ and $\varepsilon_{0}^{z}$ decrease with increasing temperatures due to a stiffing in the phonon frequencies of $A_{2u}$ and $E_{u}^{1}$ modes. The MD well
reproduced experimental values for $\varepsilon_{0}^{x}$. However, $\varepsilon_{0}^{z}$  is overestimated because the LDA pseudo-potential employed in this work underestimated $\omega_{\mathrm{TO}}$ of $A_{2u}$.
\section{Temperature-dependent BEC from MD and QHA}\label{sec:a3}
\begin{figure}[htbp]
	\centering
	\includegraphics[width=1\linewidth]{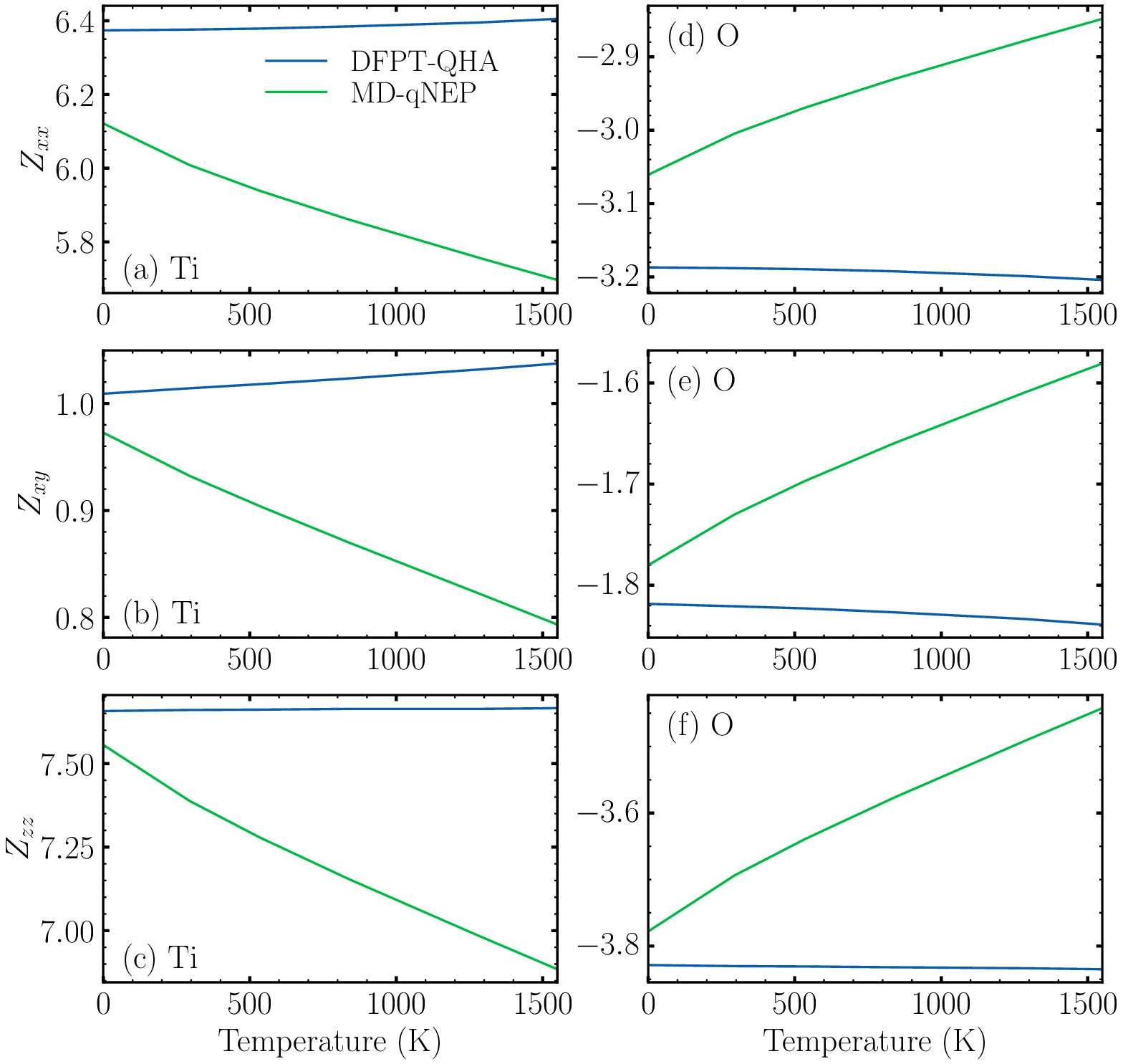}
	\caption{Temperature-dependent BEC from MD-qNEP and DFPT-QHA. Panels (a)-(c) show the Ti
		components $Z_{xx}$, $Z_{xy}$, and $Z_{zz}$,
		respectively, while panels (d)-(f) show the
		corresponding O components.}
	\label{fig:bect}
\end{figure}

We obtained the lattice constants at different temperatures from MD simulations, and then derived the Born effective charges at the corresponding temperatures using DFPT within the quasi-harmonic approximation (QHA), as shown in Fig.~\ref{fig:bect}. These results are further compared with the mean dynamic Born effective charges obtained directly from the MD simulations.
\section*{REFERENCES}
%

\end{document}